\documentclass[lettersize,journal]{IEEEtran}
\usepackage{amsmath,amsfonts}
\usepackage{algorithmic}
\usepackage{algorithm}
\usepackage{array}
\usepackage[caption=false,font=normalsize,labelfont=sf,textfont=sf]{subfig}
\usepackage{textcomp}
\usepackage{stfloats}
\usepackage{url}
\usepackage{verbatim}
\usepackage{graphicx}
\usepackage{cite}

\usepackage{color}

\usepackage{amsmath}
\usepackage{graphicx}
\begin{document}

\title{Blockchain-Aided Wireless Federated Learning: Resource Allocation and Client Scheduling}

\author{
Jun Li, 
Weiwei Zhang, 
Kang Wei, 
Guangji Chen, 
Feng Shu, 
Wen Chen, 
and Shi Jin

\thanks{(\emph{Corresponding author: Kang Wei}.)

Jun Li, Weiwei Zhang, and Guangji Chen are with the School of Electronic and Optical Engineering, Nanjing University of Science and Technology, Nanjing 210094, China(e-mail: \{jun.li, wwzhang, guangjichen\}@njust.edu.cn)

Kang Wei is with the Department of Computing, The Hong Kong Polytechnic University, Hong Kong 100872, China (e-mail: kangwei@polyu.edu.hk)

Feng Shu is with School of Information and Communication Engineering, Hainan University, Haikou 570228, China (e-mail: shufeng@163.com)

Wen Chen is with Department of Electronic Engineering, Shanghai Jiao Tong University, Shanghai 200240, China (e-mail: wenchen@sjtu.edu.cn)

Shi Jin is with the National Mobile Communication Research Laboratory, Southeast University, Nanjing 210096, China (e-mail: jinshi@seu.edu.cn)
}

}

\markboth{}
{Shell \MakeLowercase{\textit{Li et al.}}: BLOCKCHAIN-AIDED WIRELESS FEDERATED LEARNING: RESOURCE ALLOCATION AND CLIENT SCHEDULNIG}


\maketitle

\begin{abstract}
Federated learning (FL) based on the centralized design faces both challenges regarding the trust issue and a single point of failure. 
To alleviate these issues, blockchain-aided decentralized FL (BDFL) introduces the decentralized network architecture into the FL training process, which can effectively overcome the defects of centralized architecture. 
However, deploying BDFL in wireless networks usually encounters challenges such as limited bandwidth, computing power, and energy consumption. 
Driven by these considerations, a dynamic stochastic optimization problem is formulated to minimize the average training delay by jointly optimizing the resource allocation and client selection under the constraints of limited energy budget and client participation. 
We solve the long-term mixed integer non-linear programming problem by employing the tool of Lyapunov optimization and thereby propose the dynamic resource allocation and client scheduling BDFL (DRC-BDFL) algorithm.
Furthermore, we analyze the learning performance of DRC-BDFL and derive an upper bound for convergence regarding the global loss function. 
Extensive experiments conducted on SVHN and CIFAR-10 datasets demonstrate that DRC-BDFL achieves comparable accuracy to baseline algorithms while significantly reducing the training delay by 9.24$\%$ and 12.47$\%$, respectively. 
\end{abstract}

\begin{IEEEkeywords}
Federated learning, decentralized network, blockchain, resource allocation, client scheduling, Lyapunov optimization.
\end{IEEEkeywords}

\section{Introduction}
\IEEEPARstart{W}{ith} the rapid development of the Internet of Things (IoT), data from smart terminal devices are experiencing an unprecedented explosion in scale~\cite{yang2022federated,chen2020joint}. 
Machine learning (ML) technology, owing to its strong data mining and representation capabilities, has garnered extensive research and is widely applied across various domains.
However, user data is typically distributed across different terminals, posing a challenge for the traditional cloud-computing-based centralized ML architecture. 
This architecture involves transmitting data to the cloud for model training, which raises concerns about information security.
Therefore, research on ML security protection schemes is an indispensable foundation of big data applications, given the urgency and necessity of ensuring data security.
To tackle this challenge, Google proposed a distributed ML paradigm~\cite{liu2020Uncoordinated} called federated learning (FL), which avoids data transmission and conducts model training in a distributed manner. FL enables clients to train models directly on local data and upload the training results to a central server~\cite{Wei2021Low}, which realizes ML model training by fusing local models uploaded by multiple clients.

However, the conventional FL architecture relies on a secure and trusted central server to publish global models, receive client-uploaded local models, and complete model aggregation. 
This architecture suffers from the following drawbacks: 
(a) Central dependency, the central server is the central node, and there exists the issue of a single point of failure; 
(b) Auditing deficiencies, it becomes challenging to trace back to the origins if participants engage in deceptive training or upload malicious models.
To overcome these issues, relevant works have proposed the blockchain-aided decentralized FL (BDFL) architecture~\cite{Deng2022Blockchain} and applied it to the construction of ML models.
Benefited by the unique consensus mechanism and decentralized structure of the blockchain, this architecture is capable of improving the security and reliability of the FL system~\cite{Zhu2021P}. 
First, BDFL solves the issue of a single point of failure caused by the central server via peer-to-peer (P2P) communication, thereby improving the scalability of the system substantially~\cite{Tang2022G, Kalra2023D}. 
Second, blockchain serves as an immutable and traceable database. By utilizing blockchain, the BDFL system enables verification of the model's source and records the entire training history of the system, making it convenient to trace back to the source of deceptive training or malicious model uploading. 
However, in the BDFL architecture, clients need to complete both the local training and mining at the same time, which leads to resource-limited problems in terms of communication bandwidth, arithmetic power, and energy resource~\cite{Nguyen2021FL}.

In response to the aforementioned challenges, it is crucial for the BDFL system to further optimize performance metrics such as learning performance, training latency, and energy consumption under limited resources~\cite{yang2020energy, Deng2022Blockchain}.
From the theoretical aspects, an upper bound on the convergence rate based on the stochastic gradient descent (SGD) algorithm is usually derived to characterize the training performance of FL.
Based on convergence analysis, previous studies have formulated an optimization problem to improve ML learning performance while considering constraints on training delay and energy consumption, thereby designing an efficient BDFL system~\cite{Deng2022Blockchain, Nguyen2022La}. 
However, traditional resource optimization algorithms for BDFL rely on the theoretical upper bound of convergence rate, in which the practical performance is largely influenced by the parameters related to the loss function characteristics (such as Lipschitz smooth coefficient), 

In this paper, we focus on the investigation of minimizing system latency under limited resource constraints in the wireless BDFL network. Specifically, we aim to enhance the learning performance and operational efficiency of FL, by jointly optimizing the computation, energy resource, and client scheduling. 
The main contributions of this paper are summarized as follows:
\begin{itemize}
\item[$\bullet$] With the assistance of blockchain technology, we present a BDFL framework where clients perform local training and mining operations without third-party intervention. 
Subsequently, we derive the participation rate of clients according to different data distributions on the client side. 
Within the BDFL framework, we formulate a mixed-integer nonlinear programming problem for dynamic resource allocation and client scheduling aimed at minimizing average training latency under the constraints of communication topology connectivity, the number of training clients, the participation rate of clients and energy consumption. 
\item[$\bullet$] We leverage the Lyapunov optimization method to convert the long-term optimization problem into a deterministic optimization problem in each communication round. 
Specifically, we convert the long-term time average (LTA) training participation rate constraints into a penalty term incorporated into the objective function.
Based on this transformation, we propose the dynamic resource allocation and client scheduling BDFL (DRC-BDFL) algorithm via solving this mixed-integer nonlinear programming problem through iterative alternating optimization methods.
\item[$\bullet$] Our theoretical analysis unveils that the convergence rate of the proposed algorithm reaches $\mathcal{O}(\frac{1}{T})$ and we obtain the convergence upper bound related to local updating frequency and participation rate of clients. 
\item[$\bullet$] The performance of the proposed algorithm is evaluated through extensive simulation experiments. 
The evaluation results demonstrate that the proposed algorithm can effectively reduce the average training latency while achieving the same test accuracy compared to the random scheduling, round robin, and channel state-based scheduling algorithms.
\end{itemize}

\section{Related Work}
\subsection{Blockchain-aided Decentralized Federated Learning}
Traditional FL relies on a central server for client scheduling and global model aggregation, leading to issues such as central dependency and a single point of failure. 
As a decentralized paradigm and security technology, blockchain has been introduced into FL to mitigate its aforementioned limitations in recent research. 
The work in~\cite{bdfl1} proposed a blockchain-based fully decentralized peer-to-peer framework for FL. It took blockchain as the foundation, leveraging the proposed voting mechanism and a two-layer scoring mechanism to coordinate FL among participants without mutual trust, addressing issues of poisoning attacks, data leakage, and high communication costs.
Based on blockchain architecture,~\cite{Feng2021Block} proposed a decentralized horizontal FL framework to address cross-domain UAV authentication and global model updates without relying on a centralized server, while preventing privacy leakage and poisoning attacks.
By integrating blockchain into the FL framework, this DFL paradigm enhances security, resilience, and scalability, making it a promising approach for collaborative model training in distributed environments. 

Previous works focused on the design of the BDFL framework have demonstrated enhanced robustness and security of FL.
For instance, Moudoud et al.~\cite{M2022Toward} presented a DFL framework empowered by blockchain for security attack protection in IoT systems and devised a solution to minimize communication overheads by addressing the resource allocation problem under bandwidth constraints.
The work in~\cite{Warnat2021Swarm} combined edge computing with a blockchain-based P2P network to replace the central server in traditional FL, thereby enhancing system privacy.
The proposed blockchain-based DFL system in~\cite{Dir2022B} introduced a reward mechanism to ensure secure, traceable, and decentralized model aggregation updates. 
To ensure the credibility and reliability of participating devices in training, some research work leveraged blockchain to construct trust mechanisms to address the issue of participating devices in DFL. 
For example, the work in~\cite{Fan2023BB} designed an FL authentication framework by combining directed acyclic graph blockchain and accumulator, and proposed a digital signature algorithm supporting batch verification to enable anonymous interactions among nodes, which helps improve system security.
Che et al.~\cite{Che2022Adecen} proposed a serverless FL framework based on a committee mechanism, where the committee system was established to score and aggregate local gradients of selected committee members through client selection strategies and replace committee members by election strategies to filter uploaded gradient.

To improve the operational efficiency of BDFL, 
Kim et al.~\cite{Kim2019B} introduced a BDFL architecture to verify uploaded model parameters and investigate system performance metrics, e.g., learning latency and block generation rate.
In ~\cite{Pham2022BFL}, the work utilized sharding techniques to enhance the scalability of the blockchain-based federated edge learning framework. This approach not only reduces the execution time required for FL task training but also improves transaction throughput on the main chain.
However, the aforementioned works inevitably introduced a third-party blockchain network to store and verify local models, which may pose the risk of privacy leakage.
In contrast to existing research relying on the third-party blockchain network for decentralized global model aggregation, Li et al.~\cite{Li2021B} proposed a BDFL framework that integrated local training and mining steps at the client end. Firstly, without any intervention from the third-party blockchain network, the proposed BDFL framework strengthens privacy protection by preserving local models among participants. Secondly, by integrating local model training and mining at the client end, the BDFL framework improves client participation incentives and enhances the efficiency of training models, which facilitates the provision of robust global models in FL.
Based on this architecture, we formulate a stochastic dynamic optimization problem to minimize training latency under the limited energy consumption and communication resources.

\subsection{Resource Scheduling and Optimization for Decentralized Federated Learning}
In the DFL architecture, the traditional client-server-client communication mode is replaced by client-to-client communication for propagating information\cite{chen2020joint}. 
This not only addresses the issue of a single point of failure associated with the central server but also upgrades the traditional serial communication mode between clients and servers to parallel model exchange between clients.
However, resource constraints of devices, in terms of energy, communication bandwidth, and computational capability, still persist in the architecture of DFL. 
To this end, effective resource optimization plays a crucial role in further enhancing the overall efficiency of DFL operations\cite{chen2021distributed}. 

To mitigate communication overhead, related works focused on designing more efficient communication protocols to improve the communication efficiency of the system.
The work in~\cite{Kalra2023D} proposed a communication-efficient solution called ProxyFL for DFL. Regarding this scheme, each client maintains a private model and a shared proxy model, which allows clients to use the proxy model for more efficient information exchange instead of relying on the central server. 
Liao et al.~\cite{Liao2022A} introduced an adaptive control method for DFL, namely FedHP, which adaptively determines the local update frequency and constructs the network topology, leading to an improved convergence speed and higher model accuracy in heterogeneous scenarios.
In complex heterogeneous scenarios, Chen et al.~\cite{M2023Enhance} proposed a priority-based algorithm in an asynchronous DFL system to dynamically select neighbors for each edge node, which achieves a balance between communication cost and model performance.

In addition to the above issues, designing effective optimization algorithms and learning strategies is also essential in DFL. Some related works proposed resource allocation strategies based on dynamic programming, reinforcement learning, and other methods to improve learning performance and efficiency.
In~\cite{Deng2022Blockchain}, the work proposed a novel BDFL framework and formulated an optimization problem to maximize the long-term average training data size for better FL learning performance under limited energy consumption.
Nguyen et al.~\cite{Nguyen2022La} studied the latency minimization problem of BDFL in multi-server edge computing, which introduced asynchronous updates and fault-tolerant mechanisms for device failures or offline issues. It presented a novel deep reinforcement learning scheme with a parameterized advantage actor critic algorithm to jointly optimize data offloading decisions, transmit power of client devices, channel bandwidth for data offloading, and computational power allocation.
The work in~\cite{AI2023Decen} designed a clustering topology where clusters are connected through bridging devices to facilitate the decentralized propagation of aggregated models across different clusters.
Additionally, they formulated a joint optimization problem to schedule local clients and allocate computation frequency while taking into account limitations on total energy consumption and convergence speed of devices.
Compared to existing works, this paper proposes a BDFL architecture that considers multiple resource-constrained scenarios such as client-side computation resource, communication resource, and energy resource. 
Under the constraints of energy supply, we formulate a stochastic optimization problem of dynamic resource allocation and client scheduling, which minimizes the average training delay by dynamically optimizing computation frequency, mining frequency, and client scheduling.

\section{System Model and Problem Formulation}
\begin{figure*}[!t]
\centering
\includegraphics[width=5.5in]{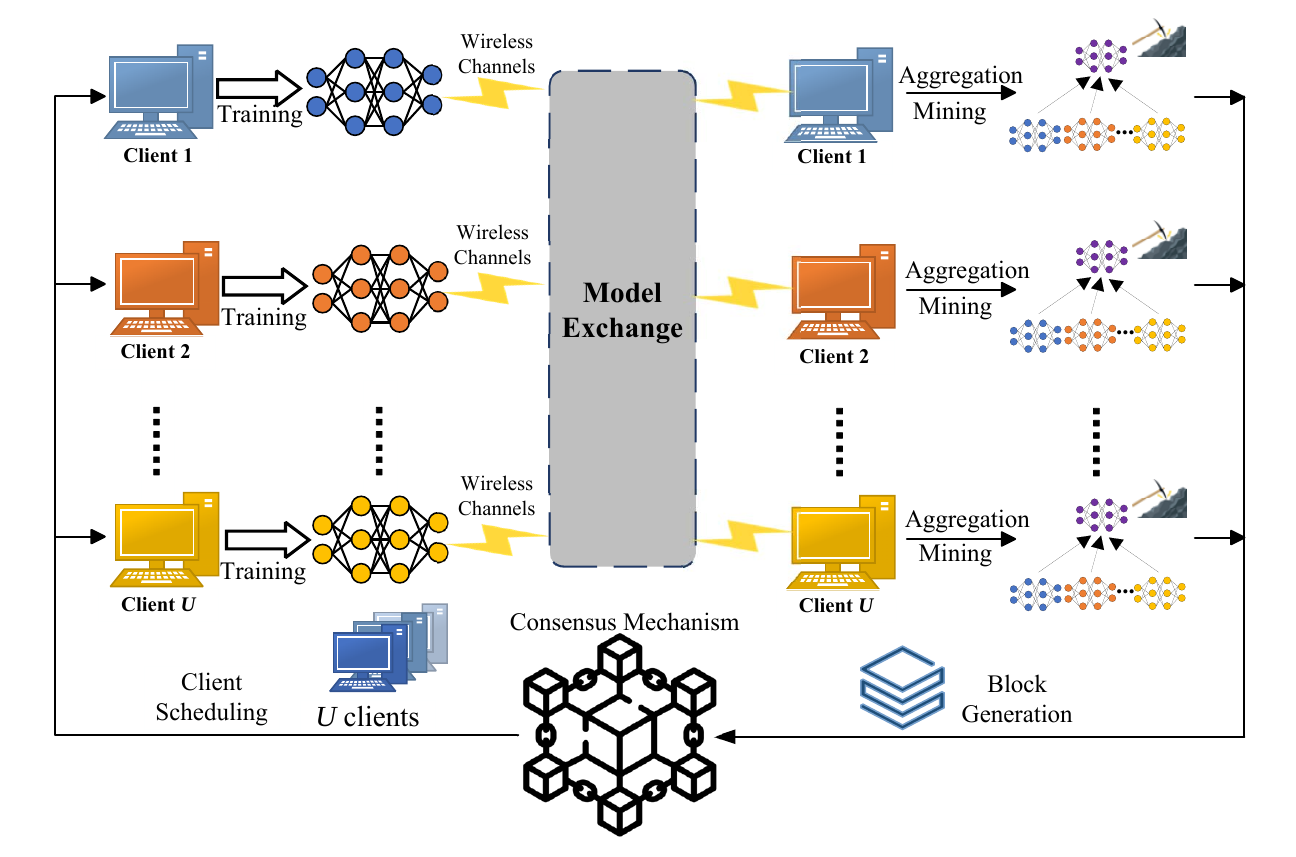}
\caption{Blockchain-aided decentralized federated learning framework}
\label{system_model}
\end{figure*}

Fig.~\ref{system_model} illustrates a wireless BDFL system consisting of \emph{U} clients. Each client can select an appropriate access point (AP), and these APs are interconnected through the wireless network. 
It should be noted that every client in the system possesses the capability for local model training and mining. 
To enable mining and local computation, we assumed that each client possesses sufficient storage resources~\cite{chen2021distributed}.
Let $\mathcal{U} = \left\{1,..., i,..., U \right\}$ denote the index set of clients, and each client has a local dataset $D_i$. Additionally, $\mathcal{T} = \left\{1,...,t,..., T \right\}$ represents the set of communication rounds.

\subsection{Federated Learning}
Unlike the existing DFL networks, the underlying FL architecture proposed in this paper incorporates blockchain to replace the central server. In this architecture, each client engages in local model training under the constraint of data distribution and conducts model transmission and mining operations. 
Considering the synchronous property of communication operations in each round of FL, the proposed model in this paper operates in the following steps in each communication round:

\textbf{(1) Client Scheduling and Local Model Training}.
The model is updated by using the mini-batch SGD algorithm, with each client performing the local update process.
Client scheduling takes place at the beginning of each communication round. For client \emph{i}, based on a mini-batch $\xi_i$ sampled from its local dataset $D_i$, the local model $\boldsymbol{w}_i(t)$ is updated in the \emph{t}-th communication round according to the gradient update rules of the local loss function. A total of \emph{H} times of local parameter updates are performed. For each training client, the update rule for the \emph{h}-th local iteration is
\begin{equation}\label{1}
\boldsymbol{w}_i^{h}(t) = \boldsymbol{w}_i^{h-1}(t) - \eta \nabla F_i\Big(\boldsymbol{w}_i^{h-1}(t)\Big) ,
\end{equation}
where $\eta$ represents the local learning rate for each client, $\boldsymbol{w}_i^{h}(t)$ denotes the local model of client \emph{i} after the \emph{h}-th local iteration in the \emph{t}-th communication round , and $\nabla F_i\Big(\boldsymbol{w}_i^{h-1}(t)\Big)$ represents the gradient. 

By denoting the \emph{h}-th local update of client \emph{i} as $g_i^h(t) = \sum\limits^{H-1}\limits_{h=0} \nabla F_i\Big(\boldsymbol{w}_i^{h}(t)\Big)$, the local update rule for client \emph{i} is given by
\begin{equation}\label{2}
    \boldsymbol{w}_i(t) = \boldsymbol{w}_i(t-1) - \eta g_i^h(t-1) .
\end{equation}

\textbf{(2) Local Model Transmission and Cross-verification}. The training clients encrypt their local models with unique digital signatures. 
Subsequently, these encrypted parameters are uploaded to the APs through wireless channels. Then, the APs exchange the received local models with other clients in the system. 
All clients in the system verify the received local models through digital signature validation and store the verified local models locally.

\textbf{(3) Global Model Aggregation and Mining}. When each client in the system receives models from other training clients, global aggregation is performed by calculating the weighted average of the local models of all training clients:
\begin{equation}\label{3}
    \boldsymbol{W}(t) = \frac{\sum\limits_{i \in{\mathbf{S}(t)}} \vert D_i\vert \boldsymbol{w}_i(t)}{\sum_{i\in\mathbf{S}(t)}\vert D_{i}\vert} ,
\end{equation}
where $\mathbf{S}(t)$ represents the set of training clients in each round and $ \sum\limits_{i\in{\mathbf{S}(t)}} \vert D_i\vert$ denotes the total size of the training dataset.

The aggregated global model is then added to the candidate block, and a random number is added to the block header. Clients compete to hash the new block again until its hash is lower than the target hash set by the block generation difficulty. 
The first client to successfully find the valid hash becomes the mining winner and is authorized to add its candidate block to the blockchain.

\textbf{(4) Block Verification and Global Model Update}. 
The mining winner propagates the new block to the entire network through the APs. 
After receiving the new block from the mining winners, the other clients verify the validity of the new block by comparing the global model in the new block with their locally aggregated models. 
If the new block is validated by the majority of clients, it is appended to the blockchain. Finally, each client updates its local model with the global model of the new block.

\subsection{Network Model}
The P2P network topology during the \emph{t}-th communication round can be represented as a connected undirected graph $\mathcal{G}(t)=(\mathcal{V},\mathcal{E}(t))$, where $\mathcal{V}$ represents the set of clients and $\mathcal{E}(t)$ represents the set of network links between clients in the \emph{t}-th communication round. The P2P network topology for the \emph{t}-th round can be represented by a symmetric adjacency matrix:
\begin{equation}\label{4}
    \textbf{A}(t)=\left\{ a_{i,j}(t) \in\left\{ 0,1 \right\}, 1 \leq i,j \leq U \right\} , 
\end{equation}
where $a_{i,j}(t)=1$ if $e_{i,j}(t) \in \mathcal{E}(t)$, otherwise 0. 

The set of training clients $\mathbf{S}(t)$ in each round can be obtained from the adjacency matrix $\textbf{A}(t)$ of each round. The set of neighbors of training client \emph{i} in the \emph{t}-th communication round is represented as $\mathcal{N}_i(t)$, which satisfies $\left|\mathcal{N}_i(t)\right|=\sum_{j\in{\mathcal{N}_i(t)}}a_{i,j}(t)$.
The degree matrix $\textbf{D}(t) = \left\{ d_{i,j}(t) , 1 \leq i,j \leq U \right\}$ is a diagonal matrix, where $d_{i,j}(t)=\left|\mathcal{N}_i(t)\right|$. Therefore, the Laplacian matrix is given by
\begin{equation}\label{5}
    \textbf{L}(t) = \textbf{D}(t) - \textbf{A}(t).
\end{equation}

According to the spectral graph theory~\cite{D2012spectral}, it is known that $\lambda_2( \textbf{L}(t))>0$ if and only if the network topology is connected, where $\lambda_n(\textbf{L}(t))>0$ denotes the \emph{n-th} smallest eigenvalue of matrix $ \textbf{L}(t)$.

\subsection{Computation Model}
Assuming that each client is equipped with a central processing unit (CPU) for local model training, the computational capability of each client's device can be measured by the CPU frequency. Specifically, we denote $f_i(t)$ as the CPU frequency of client \emph{i} in cycles per second, and $\phi_i$ as the number of CPU cycles required for one sample point to perform the forward-backward propagation algorithm. Since CPUs operate serially, the latency of local model training is given by
\begin{equation}\label{6}
d_i^{\text{cp}}(t)=\frac{\phi_i H \left| D_i\right|}{f_i(t)} ,
\end{equation}
where $H$ represents the number of iterations for local model training. Eq.~(\ref{6}) demonstrates a trade-off between the local training latency and learning performance. 
As the number of local iterations increases, the training delay also increases.
In the \emph{t}-th communication round, the CPU energy consumption of client \emph{i} during local model training~\cite{YapengZhao_FL_IRS23, JSTSP23_IRS_Computing} is defined as
\begin{equation}\label{7}
E_i^{\text{cp}}(t) = \frac{{\chi_i} \phi_i H \left| D_i\right| \left(f_i(t)\right)^2 }{2} ,
\end{equation}
where $\chi_i/2$ is the effective switched capacitance that depends on the chip architecture~\cite{Chen2022IRS,Deng2022Wire}.

\subsection{Transmission Model}
In the wireless BDFL system, an AP serves as a wireless router for data exchange between different clients. 
The uplink channel power gain from client \emph{i} to the AP is modeled as $h_i(t) = h_0 \rho_i(t)(d_0 / d_i)^v$, where $h_0$ is the path loss constant, $\rho_i(t)$ represents the small-scale fading channel power gain from client \emph{i} to the AP, and $(d_0 / d_i)^v$ denotes the large-scale path loss with \emph{v} being the path loss factor~\cite{chen2022irs1,chen2022active}. 
Therefore, the uplink data rate from client \emph{i} to the AP in the \emph{t}-th communication round can be expressed as follows
\begin{equation}\label{8}
R_i^{\text{up}}(t)=B \log_{2}{(1+\frac{P_i(t)h_i(t)}{B N_0})} ,
\end{equation}
where $B$ is the system bandwidth, $P_i(t)$ is the transmit power of client \emph{i}, and $N_0$ is the noise power spectral density.

From Eq.~(\ref{8}), it can be deduced that the transmission delay of client \emph{i} over the uplink in the t-th communication round is
\begin{equation}\label{9}
d_i^{\text{up}}(t)=\frac{m_i}{R_i^{\text{up}}(t)} ,
\end{equation}
where $m_i$ is the size of the local model of client \emph{i}, i.e., the number of bits required for client \emph{i} to transmit the local model over the wireless channel.

The energy consumption of client \emph{i} for uploading local model in the \emph{t}-th communication round can be expressed as
\begin{equation}\label{10}
E_i^{\text{up}}(t) = P_i(t) d_i^{\text{up}}(t).
\end{equation}

\subsection{Mining Model}
The mining process under the Proof of Work (PoW) consensus mechanism follows the homogeneous Poisson process. In each communication round, the mining time $d^{\text{bloc}}(t)$ is an independently and identically distributed exponential random variable with a mean of $\theta(t) = \frac{\alpha}{\sum_{i\in \mathcal{U}} f_i^{\text{bloc}}(t)}$, where $\alpha$ represents the block generation difficulty and $f_i^{\text{bloc}}(t)$ denotes the computation frequency of client \emph{i} in the \emph{t}-th round of mining. Therefore, the cumulative distribution function of the mining time in the \emph{t}-th communication round can be expressed as $\textbf{Pr}(d^{\text{bloc}}(t)<d) = 1 - e^{-\frac{d}{\theta(t)}}$. Let $p_0 = \textbf{Pr}(d^{\text{bloc}}(t)<d)$ and then the mining time in the \emph{t}-th communication round is given by
\begin{equation}\label{11}
d^{\text{bloc}}(t)= - \frac{\alpha  \ln (1-p_0)}{\sum_{i\in \mathcal{U}} f_i^{\text{bloc}}(t)} .
\end{equation}

Accordingly, the energy consumption for mining by client \emph{i} in the \emph{t}-th communication round can be expressed as
\begin{equation}\label{12}
E_i^{\text{bloc}}(t) = \frac{\chi_i d^{\text{bloc}}(t) {\left(f_i^{\text{bloc}}(t)\right)^3}}{2}. 
\end{equation}

\subsection{Problem Formulation}
Note that the waiting time for each client to collect all local models depends on the last client who completes model training and transmission. Assuming that the downlink transmission time of client parameters can be neglected compared with the overall delay~\cite{Tran2019downlink}, the latency overhead for client \emph{i} in the \emph{t}-th communication round consists of three main components, i.e.,  local model training delay, model transmission delay, and mining delay. Therefore, the total delay for each communication round is given by
\begin{equation}\label{13}
d(t) = \max\limits_{ i\in {\mathbf{S}(t)}}\left\{d_i^{\text{up}}(t) + d_i^{\text{cp}}(t)\right\} + d^{\text{bloc}}(t) . 
\end{equation}

The total energy consumption for client \emph{i} in the \emph{t}-th communication round can be represented as
\begin{equation}\label{14}
E_i(t) = E_i^{\text{up}}(t) + E_i^{\text{cp}}(t) + E_i^{\text{bloc}}(t)  .
\end{equation}

To limit client participation, we calculate the theoretical participation rate for each communication round based on data distribution. 
In particular, we first obtain the number of labels in each client's local dataset by dividing the dataset according to the distribution of their local data.  
Subsequently, we quantify the difference between the number of labels in the client's local dataset and the global proportion as a score. 
To calculate the value of $\beta_i$ for each client, we divide the sum of accumulated scores by the size of the client's dataset. 
At last, we normalize and standardize $\beta_i$ to obtain the final theoretical participation rate of each client throughout the training process.
Additionally, we define the vector $\Gamma_i(t)$ indicates whether client \emph{i} is selected to participate in the training in the \emph{t}-th communication round, which is derived from the adjacency matrix $\textbf{A}(t)$ updated in each round. 
Specifically, $\Gamma_i(t)$ takes a value of 1 if client \emph{i} is selected and 0 otherwise. 

By minimizing the average training latency of FL with the individual client's energy consumption and participation rate constraints, the corresponding resource allocation problem is formulated as
\begin{equation}\label{15}
\begin{aligned}
&\begin{array}{r@{\quad}l@{}l@{\quad}l}
&\textbf{P1}:	\min\limits_{\substack{\mathbf{S}(t),f_i(t),\\
f_i^{\text{bloc}}(t)}} \lim\limits_{T\to \infty}\frac{1}{T}\sum\limits_{t\in \mathcal{T}}d(t)\\
{\rm s.t.}	&\textbf{C1}: \lambda_2(\textbf{L}(t))>0, \forall t\in \mathcal{T},\\
	&\textbf{C2}:m\le \left| \mathbf{S}(t) \right|\le U, \forall t\in \mathcal{T},\\
	&\textbf{C3}:E_i(t)\le E_i^{\max}(t), \forall i \in \mathcal{U}, t \in \mathcal{T},\\
	&\textbf{C4}:\lim\limits_{T\to \infty} \frac{1}{T}\sum\limits_{t=1}^T\Gamma_i(t)\le \beta_i, \forall i \in \mathcal{U}.\\
\end{array}
\end{aligned}
\end{equation}

According to \textbf{P1}, computation resource, energy resource, as well as client scheduling are jointly optimized with topology connectivity constraints, energy consumption constraints, and LTA participation rate constraints.
It is observed that $\textbf{P1}$ is a long-term stochastic optimization problem involving discrete binary integer variables, which significantly increases the complexity of the problem. 
This high degree of coupling, combined with the stochastic and long-term nature of the problem, renders it impractical to solve directly using conventional convex optimization methods.

\section{Proposed Solution to P1}
In this section, we leverage the \textbf{Lyapunov optimization framework} to solve the stochastic optimization problem \textbf{P1}. 
To solve \textbf{P1}, we first utilize the Lyapunov optimization framework to convert the long-term fairness constraint into the queue stability constraint. 
By constructing the Lyapunov drift-plus-penalty function, we transform the long-term stochastic problem into the deterministic problem for each communication round. 
Then, we divide the problem into three sub-problems and iteratively determine the training clients $\mathbf{S}(t)$ in each round. 
Finally, we propose the DRC-BDFL algorithm to solve the joint optimization problem of resource allocation and client scheduling. 

First, we employ the Lyapunov optimization framework to introduce a virtual queue $Z_i(t)$ for each client. 
This allows the transformation of the long-term constraint problem $\textbf{P1}$ into a per-communication-round problem. 
To this end, we define the update equation for the virtual queue of each client as follows 
\begin{equation}\label{16}
Z_i(t+1)=\max\left\{Z_i(t)+\beta_i-\Gamma_i(t), 0 \right\} ,
\end{equation}
where $\Gamma_i(t)$ is the indicator of training participation of client $i$ and$\beta_i$ is the theoretical participation rate of client \emph{i} among all communication rounds under the constraint of data distribution, 

Additionally, $\Gamma_i(t)$ is derived from the adjacency matrix $\textbf{A}(t)$ updated in each round, which indicates whether client \emph{i} is selected to participate in the training in the \emph{t}-th communication round. 
Specifically, $\Gamma_i(t)$ takes a value of 1 if client \emph{i} is selected and 0 otherwise. 

\textit{Definition 1:} A discrete time queue $Q_i(t)$ is mean rate stable when $\lim\limits_{t \to \infty} E[Q_i(t)]/t=0$~\cite{Nee2022}.

By replacing the long-term fairness constraint $\textbf{C4}$ with the mean rate stable constraint, we transform $\textbf{P1}$ to
\begin{equation}\label{17}
\begin{aligned}
&\begin{array}{r@{\quad}l@{}l@{\quad}l}
&\textbf{P2}:	\min\limits_{\substack{\mathbf{S}(t),f_i(t),\\
f_i^{\text{bloc}}(t)}}\lim\limits_{T\to \infty}\frac{1}{T}\sum\limits_{t\in \mathcal{T}}d(t)\\
{\rm s.t.}	&\textbf{C1}: \lambda_2(\textbf{L}(t))>0, \forall t\in \mathcal{T},\\
	&\textbf{C2}:m\le \left| \mathbf{S}(t) \right|\le U, \forall t\in \mathcal{T},\\
	&\textbf{C3}:E_i(t)\le E_i^{\max}(t), \forall i \in \mathcal{U}, t \in \mathcal{T},\\
    &\textbf{C4}:\lim\limits_{T \to \infty}\frac{1}{T}\sum\limits^T\limits_{t=1}E[Z_i(t)]=0, \forall i \in \mathcal{U}.\\ 
\end{array}
\end{aligned}
\end{equation}

Definition 1 transforms the complex and intractable long-term constraint problem into an objective problem that ensures the stability of the average rate of the virtual queue throughout the FL process. 
Subsequently, we employ the Lyapunov optimization method to constrain the growth of the virtual queue $Z_i(t)$ while minimizing the objective in $\textbf{P2}$. 

\textit{Definition 2:} For each $Z_i(t)$, we define the Lyapunov function as
\begin{equation}\label{18}
\begin{aligned}
L\left( \theta(t) \right) = \frac{1}{2} \sum\limits_{i\in \mathcal{U}} Z_i^2(t).
\end{aligned}
\end{equation}

\textit{Definition 3:} Let $ \left\{ \boldsymbol{Z}(t)= Z_i(t), \forall i \in \mathcal{U} \right\}$ represent the set of virtual queues for all clients in the \emph{t}-th communication round. We define the conditional Lyapunov drift as
\begin{equation}\label{19}
\begin{aligned}
\Delta L = \sum\limits_{i\in \mathcal{U}} E\left\{L(\theta(t+1)) - L(\theta(t)) | \boldsymbol{Z}(t)\right\}.
\end{aligned}
\end{equation}

Starting from the virtual queue definition in Eq.~(\ref{16}), we can derive the conditional Lyapunov drift. 
The derivation process is as follows
\begin{equation}\label{20}
Z_i^2(t+1) \le Z_i^2(t)+\beta_i^2 + \Gamma_i^2(t)+2Z_i(t)\left( \beta_i -\Gamma_i(t)\right).
\end{equation}

By rearranging the terms, we obtain:
\begin{equation}\label{21}
\begin{split}
\Delta L &\le \sum\limits_{i\in \mathcal{U}} E\left\{Z_i(t)\left(\beta_i-\Gamma_i(t)\right) | \boldsymbol{Z}(t)\right\}\\
&\quad + \frac{1}{2} \sum\limits_{i\in \mathcal{U}} E\left\{ \beta_i^2 + \Gamma_i^2(t) | \boldsymbol{Z}(t) \right\}\\
&\le \sum\limits_{i\in \mathcal{U}} E\left\{Z_i(t)\left(\beta_i-\Gamma_i(t)\right) | \boldsymbol{Z}(t)\right\} + Q,
\end{split}
\end{equation}
where $ Q =\frac{1}{2} \sum\limits_{i\in \mathcal{U}} E\left\{ \beta_i^2 + 1 | \boldsymbol{Z}(t) \right\}$.

The conditional Lyapunov drift $\Delta L$ depends on the channel state of the current communication round, the client scheduling strategy, and the local computing resources. 
Observing inequality~(\ref{21}), the upper bound of $\Delta L$ can be obtained by minimizing the first term.
Minimizing $\Delta L$ helps stabilize the virtual queue to satisfy the mean rate stability constraint $\textbf{C4}$. 
We map the objective function of the problem $\textbf{P2}$ to a penalty function, and define the Lyapunov drift-plus-penalty function as follows 
\begin{equation}\label{22}
\begin{aligned}
\Delta_V(t) = E \left\{ Vd\left(t \right) + Z_i(t)\left(\beta_i-\Gamma_i(t)\right) | \boldsymbol{Z}(t)\right\},
\end{aligned}
\end{equation}
where $V$ is a non-negative coefficient that balances the Lyapunov drift function and the value of the objective function. Additionally, $d(t)$ represents the total time used for completing local training and global model updates in the \emph{t}-th round. 

To minimize the Lyapunov drift-plus-penalty function in Eq.~(\ref{22}), we employ the Opportunistically Minimizing an Expectation method described in~\cite{Zhang2022S} to transform the problem into each communication round. 
We adopt a scheduling strategy to minimize the following expression:

\begin{equation}\label{23}
\Delta_V(t)=\sum_{i\in \mathcal{U}}Z_i(t)(\beta_i-\Gamma_i(t))+V(d_i(t)+d^\text{bloc}(t)).
\end{equation}

It can be observed from Eq.~(\ref{23}) that the value of the queue length $Z_i(t)$ tends to be large when client \emph{i} remains consistently unselected even after the \emph{t}-th round, which leads to a large value of $\Delta_V(t)$. 
To minimize the objective function, it is preferable to schedule client \emph{i} for local training in the current communication round.

With the help of the Lyapunov drift-plus-penalty function, the problem can be ultimately transformed into $\textbf{P3}$ as follows
\begin{equation}\label{24}
\begin{aligned}
&\begin{array}{r@{\quad}l@{}l@{\quad}l}
&\textbf{P3}:	\min\limits_{\substack{\mathbf{S}(t),f_i(t),\\
f_i^{\text{bloc}}(t)}}\triangle_V(t)\\
{\rm s.t.}	&\textbf{C1}: \lambda_2(\textbf{L}(t))>0, \forall t\in \mathcal{T},\\
	&\textbf{C2}:m\le \left| \mathbf{S}(t) \right|\le U, \forall t\in \mathcal{T},\\
	&\textbf{C3}:E_i(t)\le E_i^{\max}(t), \forall i \in \mathcal{U}, t \in \mathcal{T}.\\
\end{array}
\end{aligned}
\end{equation}

With the help of the Lyapunov optimization framework, we already convert the long-term stochastic optimization problem \textbf{P1} into a deterministic optimization problem in each communication round. 
It is essential to highlight that this transformation is indeed necessary.

To address the sequence of deterministic combinatorial problems \textbf{P3}, we decouple the joint optimization problem $\textbf{P3}$ into three sub-problems. 
We then solve these sub-problems optimally in an alternating manner and obtain the optimal solution of $\textbf{P3}$ after convergence.  
In particular, we first initialize $\mathbf{S}(t) = \mathcal{U}$, i.e., making all clients participate in the training. 
Second, we initialize the virtual queue of each client to be $\boldsymbol{Z}(t) = 0$ and set their initial channel state. 
Subsequently, we optimize $f_i(t)$ and $f_i^{\text{bloc}}(t)$ using the Lagrange multiplier method respectively, and then solve $\mathbf{S}(t)$ by utilizing the optimized values of $f_i(t)$ and $f_i^{\text{bloc}}(t)$.
This process is repeated in each communication round until convergence is achieved. The detailed solution process is as follows.

For any given $f_i^{\text{bloc}}(t)$ and $\mathbf{S}(t)$, $f_i(t)$ can be optimized by solving the following problem.
\begin{equation}\label{25}
\begin{aligned}
&\begin{array}{r@{\quad}l@{}l@{\quad}l}
&\textbf{P4}:	\min\limits_{\substack{f_i(t)}}\triangle_V(t)\\
{\rm s.t.}	&\textbf{C3}:E_i(t)\le E_i^{\max}(t), \forall i \in \mathcal{U}, t \in \mathcal{T}.\\
\end{array}
\end{aligned}
\end{equation}

Firstly, we construct the Lagrange function
\begin{equation}\label{26}
L_1(f_i(t),\lambda_i)=\triangle_V(t)+\lambda_i(E_i(t)-E_i^{\max}(t)).
\end{equation}

Substitute the delay equation and the energy equation into Eq.~(\ref{26})
\begin{equation}\label{27}
\begin{aligned}
L_1(f_i(t),\lambda_i)&=\sum_{i\in \mathcal{U}}Z_i(t)(\beta_i-\Gamma_i(t))\\
&+V(\frac{m_i}{R_i^{\text{up}}(t)}+\frac{H \vert D_i\vert\Phi_i}{f_i(t)}+\frac{\alpha}{\sum_{i\in \mathcal{U}}f_i^{\text{bloc}}(t)})\\
&+\lambda_i(\frac{P_i(t)m_i}{R_i^{\text{up}}(t)}+\frac{\chi_i H \vert D_i\vert\Phi_i(f_i(t))^2}{2})\\
&+\lambda_i(\frac{\chi_i d^{\text{bloc}}(t)(f_i^{\text{bloc}}(t))^3}{2}-E_i^{\max}(t)).
\end{aligned}
\end{equation}

Taking partial derivatives of $f_i(t)$ and $\lambda_i$ in Eq.~(\ref{27}), we have:
\begin{align}\label{28}
\left\{
\begin{aligned}
\frac{\partial L_1}{\partial f_i(t)}&=-\frac{VH\vert D_i\vert\Phi_i}{(f_i(t))^2}+\lambda_i\chi_iH\vert D_i\vert\Phi_i f_i(t);\\
\frac{\partial L_1}{\partial \lambda_i}&=\frac{P_i(t)m_i}{R_i^{\text{up}}U(t)}+\frac{\chi_iH\vert D_i\vert\Phi_i(f_i(t))^2}{2}\\
\end{aligned}
\right.
\end{align}
\begin{equation}
\quad +\frac{\chi_i d^{\text{bloc}}(t)(f_i^{\text{bloc}}(t))^3}{2}-E_i^{\max}(t).\notag
\end{equation}

By setting $\frac{\partial L_1}{\partial f_i(t)}=0$ and $\frac{\partial L_1}{\partial \lambda_i}=0$, we can solve for $f_i(t)$ 
\begin{equation}
    f_i(t)	=\sqrt{\frac{2(E_i^{\max}(t)-\frac{P_i(t)m_i}{R_i^{\text{up}}(t)}-\frac{\chi_i d^{\text{bloc}}(t)(f_i^{\text{bloc}}(t))^3}{2})}{\chi_iH\vert D_i\vert\Phi_i}}
\end{equation}

For any given $f_i(t)$ and $\mathbf{S}(t)$, $f_i^{\text{bloc}}(t)$ can be optimized by solving the following problem.
\begin{equation}\label{30}
\begin{aligned}
&\begin{array}{r@{\quad}l@{}l@{\quad}l}
&\textbf{P5}:	\min\limits_{\substack{f_i^{\text{bloc}}(t)}}\triangle_V(t)\\
{\rm s.t.}	&\textbf{C3}:E_i(t)\le E_i^{\max}(t), \forall i \in \mathcal{U}, t \in \mathcal{T}.\\
\end{array}
\end{aligned}
\end{equation}

By employing the Lagrange multiplier method to solve $f_i^{\text{bloc}}(t)$, the solution process remains the same as that for $f_i(t)$, and ultimately $f_i^{\text{bloc}}(t)$ can be solved
\begin{equation}
\begin{aligned}
f_i^{\text{bloc}}(t)	&=\sqrt[3]{\frac{MN}{2}+ \sqrt{\left( \frac{MN}{2}\right) - \left( \frac{M}{3}\right)^3 }} \\
&+ \sqrt[3]{\frac{MN}{2} - \sqrt{\left( \frac{MN}{2}\right) - \left( \frac{M}{3}\right)^3 }},\\
\end{aligned}
\end{equation}

where $M = -\frac{2(E_i^{\max}(t)-\frac{P_i(t)m_i}{R_i^{\text{up}}(t)}-\frac{\chi_i H \left| D_i \right| \Phi_i (f_i(t))^3}{2})}{\chi_i\alpha \ln{(1-p_0)}}  $, \\
$N= \sum\limits_{\substack{\forall j \in N, \\j\neq i}} f_j^{\text{bloc}}(t)$,


For any given $f_i(t)$ and $f_i^{\text{bloc}}(t)$, $\mathbf{S}(t)$ can be optimized by solving the following problem.
\begin{equation}\label{32}
\begin{aligned}
&\begin{array}{r@{\quad}l@{}l@{\quad}l}
&\textbf{P6}:	\min\limits_{\substack{\mathbf{S}(t)}}\triangle_V(t)\\
{\rm s.t.}	&\textbf{C1}: \lambda_2( \textbf{L}(t))>0, \forall t\in \mathcal{T},\\
	&\textbf{C2}:m\le \left| \mathbf{S}(t) \right|\le U, \forall t\in \mathcal{T},\\
    &\textbf{C3}:E_i(t)\le E_i^{\max}(t), \forall i \in \mathcal{U}, t \in \mathcal{T}.\\
\end{array}
\end{aligned}
\end{equation}

According to constraint $\textbf{C2}$, the number of training clients in each round ranges from $\left[m, U \right]$, and the total delay $d_i(t)$ for each client can be solved by substituting them into the delay formula using $f_i(t)$ and $f_i^{\text{bloc}}(t)$ obtained by solving $\textbf{P4}$ and $\textbf{P5}$. 
To begin with, we sort the latency $d_i(t)$ of all clients from smallest to largest to form a sequential latency list $\left\{d^{\prime}(t) \right\}$. 
Subsequently, we generate the corresponding client index list $\left\{U^{\prime}(t) \right\}$ according to $\left\{d^{\prime}(t) \right\}$.
Then we divide the clients into ($U-m$) groups, with each group consisting of the first $m$, $m+1$, ......, $U$ clients of list $\left\{U^{\prime}(t) \right\}$, respectively. 
To minimize the objective function, each group of training clients is substituted to calculate its corresponding objective function, and the group of clients that obtains the minimum value of the objective function is chosen as $\mathbf{S}(t)$.

In each communication round, the aforementioned solving steps are iteratively performed until the set of participating clients $\mathbf{S}(t)$ stabilizes, at which point the solving process is terminated. Based on the above optimization procedure, the overall workflow of DRC-BDFL is introduced in Algorithm~\ref{alg:algorithm1}.
\begin{algorithm}
\renewcommand{\algorithmicrequire}{\textbf{Input:}}
\renewcommand{\algorithmicensure}{\textbf{ENSURE:}}
\newcommand{\LASTCON}{\item[\algorithmiclastcon]}
\newcommand{\algorithmiclastcon}{\textbf{Output:}}
\footnotesize
\caption{DRC-BDFL}
\label{alg:algorithm1}
\begin{algorithmic}[1]
    \STATE Initialize $\boldsymbol{Z}=0$, \emph{T}, \emph{V}, $\boldsymbol{f}$, $\boldsymbol{f}^{\text{bloc}}$; 
    \STATE \textbf{for \emph{t}=1 to \emph{T} at client $i\in \mathcal{U}$ do}
        \STATE \qquad\textbf{Require}: Virtual queue length $\left\{Z_i(t)\right\}$ and channel state$\left\{s_i(t)\right\}$; 
        \STATE \qquad Initialize $n$=0;
            \STATE \qquad\textbf{Repeat}    
                \STATE \qquad\qquad find $f_i^n(t)$ according to \textbf{(29)};
                \STATE \qquad\qquad find $f_i^{\text{bloc},n}(t)$ according to \textbf{(31)};
                \STATE \qquad\qquad find $\boldsymbol{S}^n(t)$ by solving \textbf{P6};
                \STATE \qquad\qquad Set $n$ $\Leftarrow$ $n+1$;
            \STATE\qquad \textbf{Until} $\boldsymbol{S}^{n+1}(t) = \boldsymbol{S}^{n}(t)$;
            \STATE\qquad \textbf{Update} $\mathbf{S}(t) \Leftarrow \boldsymbol{S}^n(t)$, $f_i(t+1) \Leftarrow f_i(t)$, $f_i^{\text{bloc}}(t+1) \Leftarrow f_i^{\text{bloc}}(t)$;
            \STATE\qquad \textbf{Update} $\left\{Z_i(t)\right\}$; 
    \STATE\textbf{end}     
    \LASTCON $\boldsymbol{f}$, $\boldsymbol{f}^{\text{bloc}}$, $\mathbf{S}(t)$.
\end{algorithmic}
\end{algorithm}

\section{Convergence And Complexity Analysis}
\subsection{Convergence Analysis}
In this section, we analyze the convergence performance of the proposed algorithm, and characterize the convergence upper bound in terms of the number of communication rounds $T$, the number of clients $U$, the number of local iterations $H$, the learning rate $\eta$, the data distribution, and the theoretical participation rate $\beta$. 
We first introduce some assumptions that are widely used in convergence analysis~\cite{Kolo2022, chen2020convergence}.

\textit{Assumption 1:} We assume the global loss function ${F}\left( \boldsymbol{w} \right)$ is $L$-smooth, namely $\forall \boldsymbol{w}(t+1),\boldsymbol{w}(t),t \in \left[ {1,T} \right]$, the objective function satisfies:
\begin{equation}\label{33} 
\begin{aligned}
&F(\boldsymbol{w}(t+1)) - F(\boldsymbol{w}(t)) \le \Big< \nabla F(\boldsymbol{w}(t)), w(t+1)- w(t) \Big> \\
&+ \frac{L}{2} \| \boldsymbol{w}(t+1)-\boldsymbol{w}(t) \|^2 .
\end{aligned}
\end{equation}

\textit{Assumption 2:} The stochastic gradient of each client is bounded.
\begin{equation}\label{34}
\begin{split}
E[\left\| \nabla F_i(\boldsymbol{w}) \right\|^2] \le G^2 ,  \forall \boldsymbol{w},i .
\end{split}
\end{equation}

\textit{Assumption 3:} The global loss function has a lower bound, which satisfies:
\begin{equation}\label{35}
\begin{split}
 F(\boldsymbol{w}(t))> -\infty .
\end{split}
\end{equation}

\textit{Lemma 1:} Let Assumption 1, Assumption 2, and Assumption 3 all hold, then the convergence bound of Algorithm 1 can be given as
\begin{equation}\label{36}
\begin{aligned}
&\frac{1}{T}\sum\limits_{t=1}^T E[\left\|F(\boldsymbol{w}(t+1))\right\|^2] 
\le \frac{2 E[F(\boldsymbol{w}(1))-F(\boldsymbol{w}^*)]}{\eta HT}\\
&+2\eta L HG^2 \sum\limits_{i=1}^U \frac{\beta_i\left|D_i\right|}{\left|D\right|} \sum\limits_{i=1}^U \frac{(1-\beta_i) \left|D_i\right| }{\left| D\right|}\\
&+\eta ULH G^2*(\eta H + 1) \sum\limits_{i=1}^U \frac{\beta_i^2\left|D_i\right|^2}{\left|D\right|^2}\\
&+UG^2\sum\limits_{i=1}^U \frac{(1-\beta_i)^2 \left|D_i\right|^2}{\left| D\right|^2}.\\
\end{aligned}
\end{equation}

\textit{Proof.} Please see \textbf{Appendix A}.

According to \textbf{Lemma 1}, the convergence rate of this algorithm can be controlled within a certain range. At the same time, with the increase of communication rounds, the cumulative average gradient value decreases as $T$ increases. As $T$ approaches infinity, the gradient approximately converges to zero. The algorithm proposed in this paper guarantees the convergence mathematically, and the convergence rate reaches the order of $\mathcal{O}(\frac{1}{T})$. 

\subsection{Complexity Analysis}
In this subsection, we analyze the computational complexity of the DRC-BDFL algorithm. 
For $T$ communication rounds of $U$ clients, the computational complexity of the outer loop can be expressed as $\mathcal{O}(T)$.
At the beginning of each communication round, the computational complexity of initializing the virtual queues and channel states is $\mathcal{O}(U)$.
As mentioned above, we decompose the mixed-integer nonlinear program problem $\textbf{P3}$ into three sub-problems $\textbf{P4}$, $\textbf{P5}$, and $\textbf{P6}$, which have $\mathcal{O}(U)$, $\mathcal{O}(U^2)$, and $\mathcal{O}(U \log U)$, respectively. 
Let $K$ represent the number of iterations required to converge the three sub-problems in the inner loop.  
In summary, under $T$ communication rounds, the overall computational complexity of the DRC-BDFL algorithm can be expressed as $\mathcal{O}(T K U^2)$.

\begin{table}[!t]
\caption{List of experimental parameters\label{tab3}}
\centering
\begin{tabular}{|c|c|c|c|}
\hline
Parameters & Values & Parameters & Values \\
\hline
$m$ & 3 & $B$ & 180 KHz\\
\hline
$\alpha$ & $3\times10^7$ & $h_0$ & $10^{-3}$\\
\hline
$d_0$ & 1 m & $d_n$ & 200 m\\
\hline
$N_0$ & $10^{-16}$ & $\phi_i$ & $5\times 10^3$\\
\hline
$m_i$& $10^6$ & $E^{\max}$ & 0.4 J\\
\hline
$D_i$& $3\times 10^{3}$& $p_0$&$10^{-10}$\\
\hline
\end{tabular}
\end{table}

\begin{figure}[!t]
\centering
\includegraphics[width=3in]{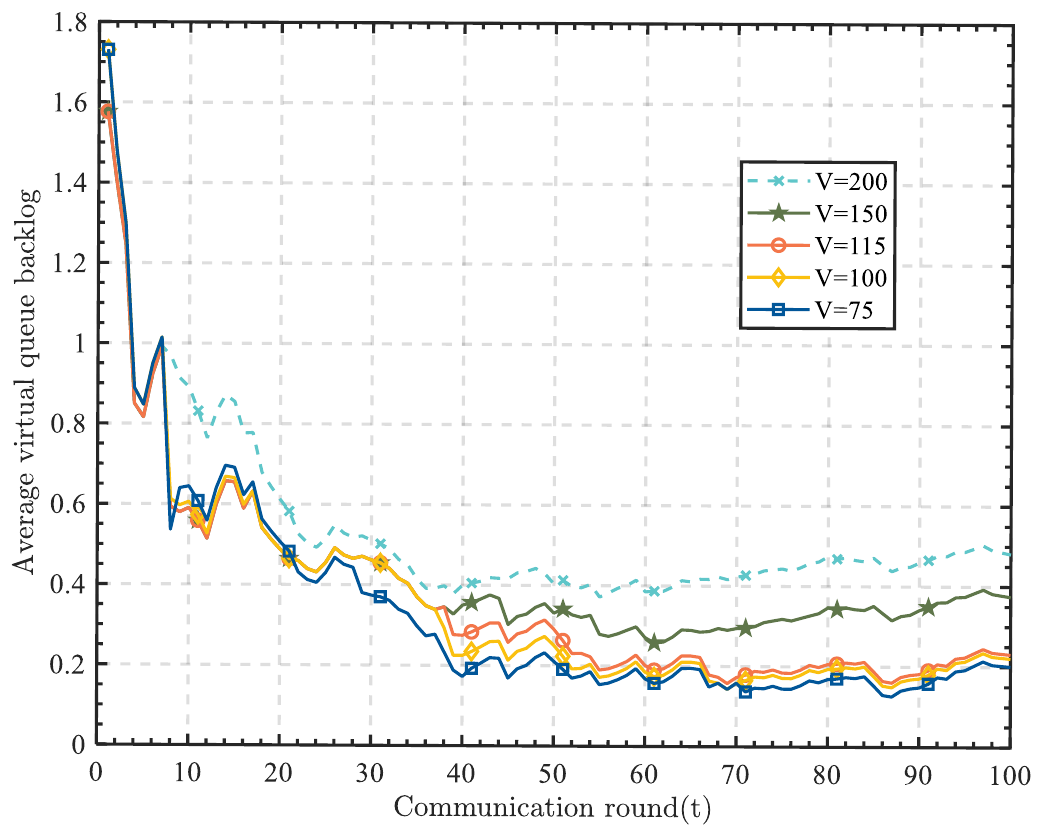}
\caption{Average virtual queue backlog with different $V$ values}
\label{differV}
\end{figure}

\section{Experimental Results}
\subsection{Experimental Setting}
\subsubsection{Datasets}
This study utilizes the SVHN dataset \cite{svhn} and CIFAR-10 dataset \cite{cifar} to conduct experiments.
The SVHN dataset, derived from Google Street View images, comprises door numbers ranging from 0 to 9. It includes a total of 600,000 color images with a resolution of 32×32, subdivided into 73,257 training samples and 26,032 test samples. 
On the other hand, the CIFAR-10 dataset encompasses 60,000 RGB three-channel color images, distributed across 10 categories with each category containing 6,000 images. The dataset is divided into 50,000 training samples and 10,000 test samples.

\subsubsection{Model}
The convolutional neural network (CNN) employed in this paper is structured as follows: It begins with two 5$\times$5 convolutional layers, the first containing 64 channels and the second comprising 128 channels. Each of these convolutional layers is activated by the ReLU function and is followed by a 3$\times$3 pooling layer. Subsequently, the network includes three fully connected layers with 2048, 384, and 192 units respectively. The first two fully connected layers are also activated by the ReLU function. The network architecture concludes with a softmax output layer.

\subsubsection{Experimental Details}
 In real-world scenarios, data consistency among different clients may not be guaranteed. In other words, disparate clients may collect highly divergent or even contradictory samples. 
To fit these conditions more accurately, the FL experiment is conducted under the non-independent and identically distributed (Non-IID) scenario. 
The two datasets are divided according to the Dirichlet distribution, guided by the parameter $\alpha$ ($\alpha > 0$), which serves as a centralizing factor controlling consistency among clients~\cite{Diri}. 
Here, a smaller Non-IID degree value (i.e., the distribution parameter $\alpha$) indicates a higher level of data heterogeneity. Conversely, as this value gets higher, the data distribution becomes more homogeneous.

\begin{figure*}[!t]
\centering
\includegraphics[width=1.0\textwidth]{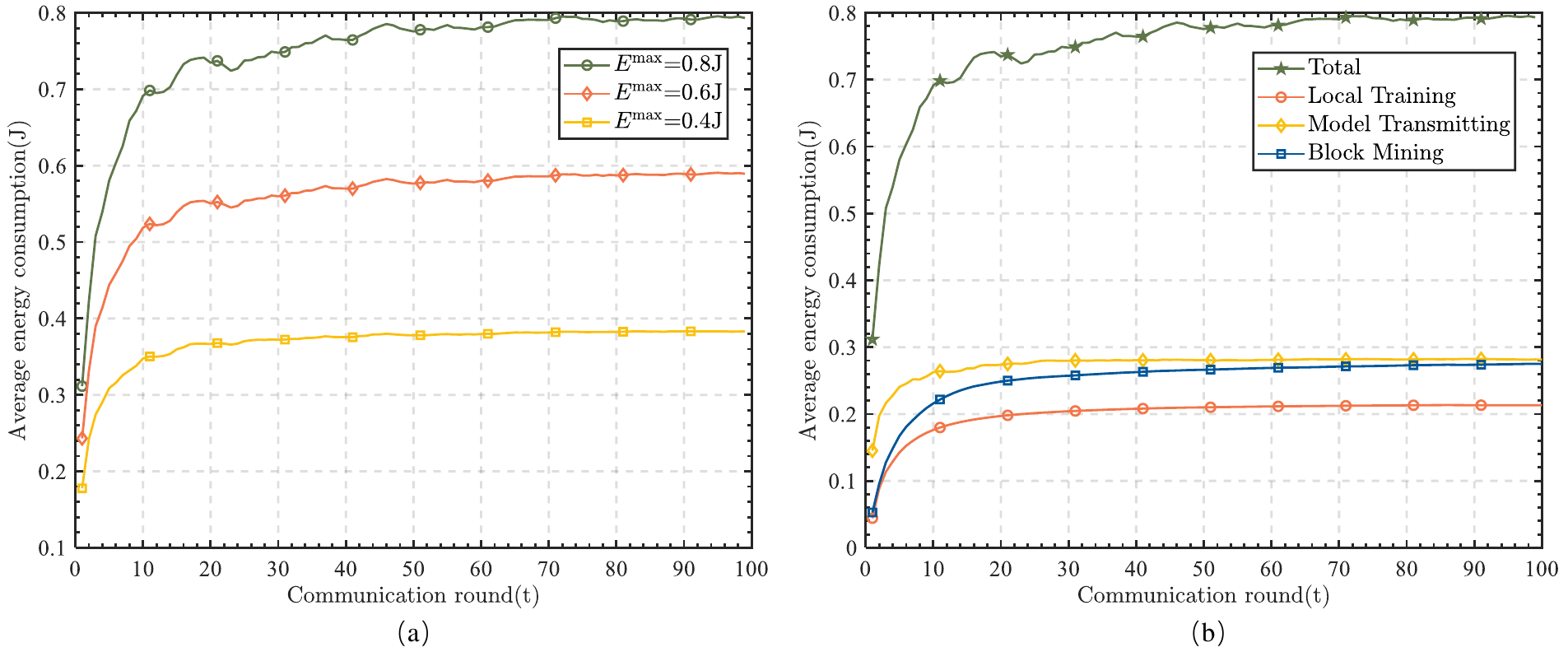}
\caption{Average energy consumption (a) different energy consumption limits; (b) energy consumption of each part ($E^{\max}=0.8\text{J}$)}
\label{differE}
\end{figure*}

In this study, a total of 8 clients are involved in the experiment. The total number of local iterations \emph{H} is set to 20, and the total number of communication rounds \emph{T} is set to 100.
For the SVHN data set, the learning rate $\eta$ is set to 0.01, and the Non-IID degree is set to 0.5; For the CIFAR-10 dataset, set the learning rate $\eta$ to 0.001 and the Non-IID degree to 0.3. 
We set the initial values of $f_i(t)$ and $f_i^{\text{bloc}}(t)$ to 1GHz and 1.5GHz, respectively.
Other specific parameter settings are shown in Table \ref{tab3}.

In order to better demonstrate the performance of DRC-BDFL, the following three baseline algorithms are used for comparison in the experiments: 
\begin{itemize}
\item[(a)] Random scheduling (abbreviated as ``Random''), which randomly selects clients to participate in the training in each round~\cite{Deng2022Blockchain,Wei2021Low}.
\item[(b)] Round robin (abbreviated as ``RR''), which cyclically selects clients to participate in the training following the client's index~\cite{hu2023s}.
\item[(c)] Client scheduling based on channel state (abbreviated as ``CB''), which selects clients with optimal channel states to participate in the model training~\cite{xia2020multi,Chu2022}.
\end{itemize}

Note that all three baseline algorithms, along with the proposed DRC-BDFL algorithm, have an equal number of training clients per communication round.

\subsection{Parameter Sensitivity Analysis}


In this section, we demonstrate the effectiveness of the proposed DRC-BDFL via extensive experimental results.

Fig.~\ref{differV} illustrates the variation of the clients' average virtual queue backlog of clients with different values of \emph{V} over communication rounds on two datasets. 
As mentioned in Section 4, the parameter \emph{V} represents a non-negative weight coefficient that balances the Lyapunov drift and training latency according to Eq.~(\ref{23}). 
Within a certain range, a larger value of \emph{V} corresponds to a higher weight on training latency, and vice versa, a smaller value corresponds to a lower weight.
the larger value of \emph{V} corresponds to the higher weight on training latency, and vice versa, the smaller it is. 
From Fig.~\ref{differV}, it can be observed that the average virtual queue backlog of clients decreases and stabilizes over time. 
Comparing the same number of communication rounds, a larger value of \emph{V} leads to a higher average virtual queue backlog, indicating a greater weight on training latency. 
In this case, the scheduling strategy tends to reduce the training latency, which results in fewer training clients (i.e., $\Gamma_i(t)$ becomes zero in Eq.~(\ref{23}), leading to an increased virtual queue backlog. 
This observation aligns with the derivation results presented in Section 4.

The curve in Fig.~\ref{differE}(a) depicts the average actual energy consumption of clients in the DRC-BDFL algorithm under different energy constraints (i.e., the maximum energy constraint $E^{\max}$ in the constraint $\textbf{C3}$).
The experimental results reveal a gradual increase in the average actual total energy consumption of clients over time, eventually converging to the predetermined maximum energy consumption $E^{\max}$. 
This indicates that as time progresses, clients make more extensive use of the energy resources in the system. 
The experiments have proved that the optimization problem solution guarantees 
the satisfaction of the given constraint $\textbf{C3}$.
Taking the example of setting $E^{\max}$ to 0.8J, Fig.~\ref{differE}(b) exhibits the variations in local training energy, model transmission energy, and mining energy, indicating that the changes in these three components of energy consumption follow the same trend as the total energy consumption. 
They all increase over time and tend to reach a maximum value.

\subsection{Performance Comparison}
\begin{figure*}[!t]
\centering
\includegraphics[width=1.0\textwidth]{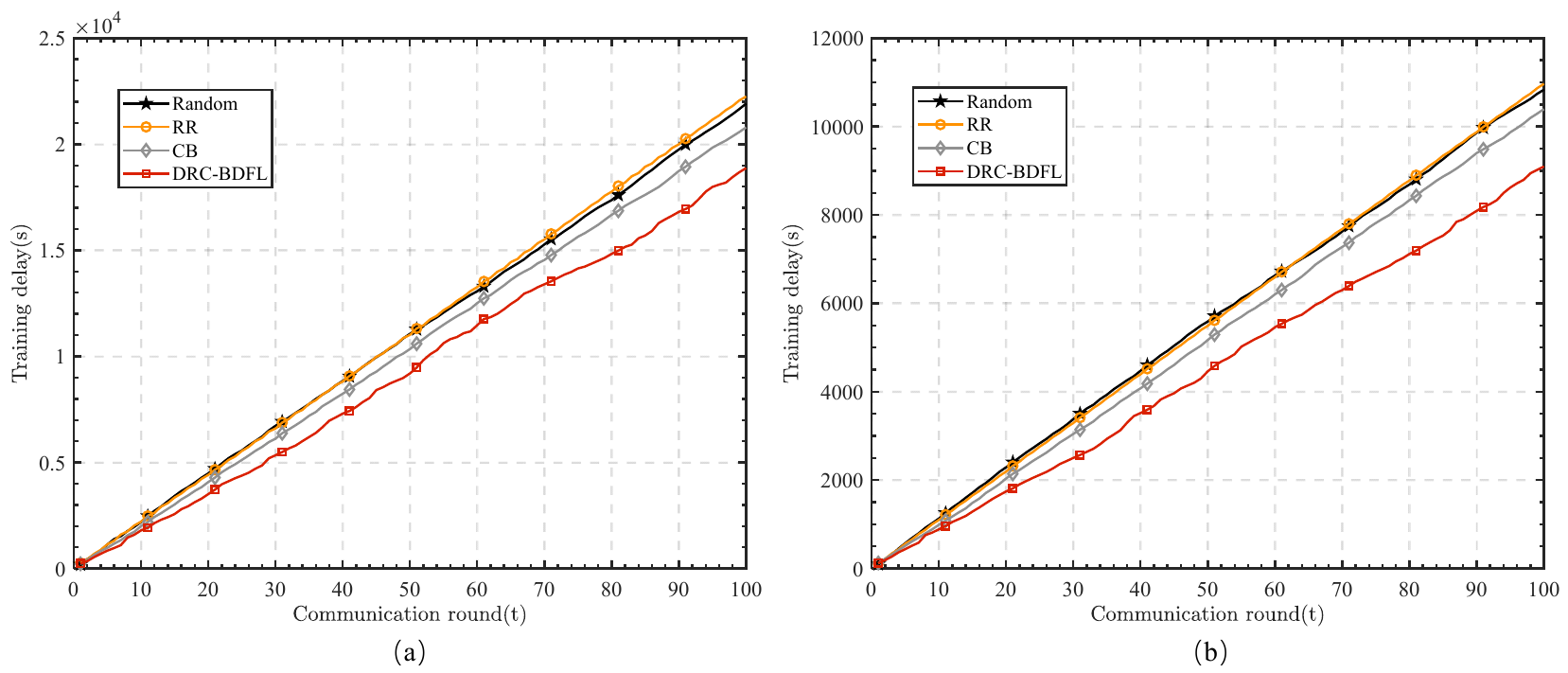}
\caption{Training delay comparison between different algorithms on two datasets. (a) SVHN;  (b) CIFAR-10}
\label{delay}
\end{figure*}
This section presents the experimental results that evaluate the training latency and test accuracy performance of the proposed algorithm DRC-BDFL. The evaluation is conducted using the SVHN and CIFAR-10 datasets.

Extensive experiments are conducted on the SVHN and CIFAR-10 datasets to compare the proposed algorithm DRC-BDFL with three baseline algorithms. 
The experimental results for the average test accuracy performance and average training latency per round are shown in Table \ref{tab4}. 
Our findings indicate that DRC-BDFL achieves significantly lower training latency than the minimum latency among the three baseline algorithms, with reductions of 9.24$\%$ and 12.47$\%$ on the SVHN and CIFAR-10 datasets, respectively.
Additionally, under the condition of achieving similar test accuracy, DRC-BDFL effectively reduces the average training latency and minimizes the communication overhead of the system.
\begin{table*}[!t]
\caption{Comparison of accuracy and training delay}
\label{tab4}
\centering
\tabcolsep 20pt
\renewcommand{\arraystretch}{1.25}
\begin{tabular}{c|c|c|c|c}
\hline
& \multicolumn{2}{c}{SVHN}&\multicolumn{2}{c}{CIFAR-10} \\ \hline
&accuracy($\%$)&training time(s) &accuracy($\%$) &training time(s)\\\hline
Random & 91.134 & 219.128 & 74.85& 108.310\\
RR & 91.563 & 222.479 &75.66& 109.658 \\
CB & 91.627& 207.962 & 75.43& 103.929\\
$\boldsymbol{{\rm Ours }}$& 91.610 & $\boldsymbol{188.750}$& 74.57& $\boldsymbol{90.964}$\\
\hline
\end{tabular}
\end{table*}

Fig.~\ref{delay} demonstrates the variation of model cumulative training delay with communication rounds for the DRC-BDFL algorithm and three baseline algorithms (Random, RR, and CB) on the SVHN and CIFAR-10 datasets, respectively. 
The number of selected clients in the three baseline algorithms is the same as that of the DRC-BDFL algorithm in the experiments.
It can be observed that the training latency of all three baseline algorithms is significantly higher than that in DRC-BDFL. 
Among the baseline algorithms, CB exhibits a slightly lower training delay than the other two baseline algorithms on both datasets due to its consideration of the client's channel state.
The proposed DRC-BDFL algorithm outperforms the baseline algorithm in terms of training delay, and the advantage is more obvious as the communication rounds increase.
This is because the client scheduling in the DRC-BDFL algorithm relies on all the relevant information considered in Section 4, and is performed by jointly optimizing the computational power, communication and energy consumption, and channel state. 
The experimental results confirm that DRC-BDFL effectively optimizes resource allocation and client scheduling while operating within constrained energy consumption during system training.

\section{Conclusion}
In this paper, we have proposed a novel BDFL architecture and investigated the problem of dynamic client scheduling and resource optimization in FL.
We have established corresponding mathematical models for the processes of model training as well as model aggregation, and formulated a joint optimization problem to minimize the training latency as the objective function for client selection and resource allocation.
Based on the Lyapunov optimization method, we have solved the long-term mixed-integer nonlinear programming problem and proposed an efficient resource optimization and client scheduling algorithm called DRC-BDFL.
Then, we have analyzed the convergence of the proposed algorithm and obtained an upper bound on the convergence. Its convergence speed reached the order of $\mathcal{O}(\frac{1}{T})$ and provided a theoretical convergence guarantee.
We have evaluated the performance of DRC-BDFL through extensive simulations and the results have demonstrated the superiority of DRC-BDFL over other baseline algorithms.

One possible future direction of BDFL architecture is to enhance privacy protection by researching stronger privacy-preserving techniques to ensure data security and client privacy~\cite{Wei2023PFL}. 
This includes developing advanced cryptographic methods, such as homomorphic encryption and secure multi-party computation, to protect data during model training and aggregation. 
Besides, integrating FL with 6G technology by combining sensing and communication functions into a unified framework is another promising research direction~\cite{Lu2024}. 
By leveraging the distributed computation and data processing capabilities of FL, cross-layer resource management methods can be designed for integrated sensing and communications (ISAC) to enhance its overall performance and resilience. Moreover, FL, as a distributed learning framework, can be utilized to address the multi-object multi-task recognition problem in ISAC systems.

{\appendix[Proof of Lemma 1]
First, we obtain the updated equation for the local model of client \emph{i} in Section II.B is
\begin{equation}\label{37}
\boldsymbol{w}_i(t+1) = \boldsymbol{w}_i(t) - \eta  \sum\limits^{H-1}\limits_{h=0} \nabla F_i(\boldsymbol{w}_i^{h}(t)).
\end{equation}

Substitute the global aggregation equation in Section 3.1 into the Eq.~(\ref{37}), and the global model update equation can be derived as
\begin{equation}\label{38}
\boldsymbol{w}(t+1) = \boldsymbol{w}(t) - \eta \sum\limits_{i=1}^U \Gamma_i(t)\frac{\left|D_i\right|}{\left|D\right|} \sum\limits^{H-1}\limits_{h=0} \nabla F_i(\boldsymbol{w}_i^{h}(t)).
\end{equation}

According to the Lipschitz smoothness in Assumption 1, substituting the gradient update equation \eqref{38} yields
\begin{equation}\label{39}
\resizebox{1.0\hsize}{!}{$
\begin{aligned}
&E[F(\boldsymbol{w}(t+1)) | \boldsymbol{w}(t)] \\
&\le F(\boldsymbol{w}(t))+\big<\nabla F(\boldsymbol{w}(t)), E[\boldsymbol{w}(t+1)-\boldsymbol{w}(t)|\boldsymbol{w}(t)]\big> \\
&+\frac{L}{2}E\Big[\left\| \boldsymbol{w}(t+1)-\boldsymbol{w}(t)\right\|^2|\boldsymbol{w}(t)\Big] \\
&\le  F(\boldsymbol{w}(t))+\frac{\eta^2 L}{2}E\Bigg[ \left\| \sum\limits_{i=1}^U \Gamma_i(t)\frac{\left|D_i\right|}{\left|D\right|} \sum\limits_{h=0}^{H-1}\nabla F_i(\boldsymbol{w}_i^h(t)) \right\|^2 \Big| \boldsymbol{w}(t)\Bigg] \\
&- \eta \Big< \nabla F(\boldsymbol{w}(t)),E[\sum\limits_{i=1}^U \Gamma_i(t)\frac{\left|D_i\right|}{\left|D\right|} \sum\limits_{h=0}^{H-1}\nabla F_i(\boldsymbol{w}_i^h(t)) | \boldsymbol{w}(t)]\Big>\\
& \le F(\boldsymbol{w}(t))\\
&- \eta \Big<\nabla F(\boldsymbol{w}(t)),\sum\limits_{i=1}^U \frac{\left|D_i\right| E[\Gamma_i(t)|\boldsymbol{w}(t)]}{\left|D\right|} \sum\limits_{h=0}^{H-1}E[\nabla F_i(\boldsymbol{w}_i^h(t)) \Big| \boldsymbol{w}(t)]\Big>\\
&+\frac{\eta^2 LUH}{2} \sum\limits_{i=1}^U \frac{\left|D_i\right|^2 (E[\Gamma_i(t)|\boldsymbol{w}(t)])^2}{\left|D\right|^2} \sum\limits_{h=0}^{H-1} E\Big[\left\| \nabla F_i(\boldsymbol{w}_i^h(t)) \right\|^2 \Big| \boldsymbol{w}(t)\Big] ,\\
\end{aligned}
$}
\end{equation}

where the last step comes from the Jensen's Inequality, $\left\| \frac{1}{N} \sum\limits_{n=1}^N x_n\right\|^2 \le \frac{1}{N} \sum\limits_{n=1}^N \left\|x_n\right\|^2$, $\left\| \sum\limits_{n=1}^N x_n\right\|^2 \le N \sum\limits_{n=1}^N \left\|x_n\right\|^2$.

According to the previous analysis, $\Gamma_i(t)$ in inequality~(\ref{39}) denotes the indicator vector of whether client \emph{i} participates in the training in the \emph{t}-th round, and $\beta_i$ is the theoretical participation rate of client \emph{i} in all communication rounds. 
Then, we have $E[\Gamma_i(t)]=E[\Gamma_i(t)|\boldsymbol{w}(t)]=\beta_i$, and inequality~(\ref{39}) can be simplified as follows:
\begin{equation}\label{40}
\resizebox{1\hsize}{!}{$
\begin{aligned}
&E[F(\boldsymbol{w}(t+1)) | \boldsymbol{w}(t)] \\
&\le F(\boldsymbol{w}(t))- \eta \Big<\nabla F(\boldsymbol{w}(t)),\sum\limits_{i=1}^U \frac{\left|D_i\right| \beta_i}{\left|D\right|} \sum\limits_{h=0}^{H-1}E[\nabla F_i(\boldsymbol{w}_i^h(t)) \Big| \boldsymbol{w}(t)]\Big>\\
&+\frac{\eta^2 LUH}{2} \sum\limits_{i=1}^U \frac{\left|D_i\right|^2 \beta_i^2}{\left|D\right|^2} \sum\limits_{h=0}^{H-1} E\Big[ \left\| \nabla F_i(\boldsymbol{w}_i^h(t))  \right\|^2 \Big| \boldsymbol{w}(t)\Big]. \\
\end{aligned}
$}
\end{equation}

By the property of conditional expectation, take the expectation for both the left and right sides of inequality~(\ref{40}).
\begin{equation}\label{41}
\resizebox{1.0\hsize}{!}{$
\begin{aligned}
&E[F(\boldsymbol{w}(t+1))]\\
& \le E[F(\boldsymbol{w}(t))] - \underbrace{ \eta \sum\limits_{h=0}^{H-1}E\Big[\Big<\nabla F(\boldsymbol{w}(t)),\sum\limits_{i=1}^U \frac{\left|D_i\right| \beta_i}{\left|D\right|} \nabla F_i(\boldsymbol{w}_i^h(t)) \Big>\Big]}_{A4(a)}\\
&+\underbrace{\frac{\eta^2 LUH}{2} \sum\limits_{i=1}^U \frac{\left|D_i\right|^2 \beta_i^2}{\left|D\right|^2} \sum\limits_{h=0}^{H-1} E\Big[ \left\| \nabla F_i(\boldsymbol{w}_i^h(t))  \right\|^2 \Big]}_{A4(b)} .\\
\end{aligned}
$}
\end{equation}

Constrain the second term $A4(a)$ of inequality~(\ref{41}):
\begin{equation}\label{42}
\resizebox{1.0\hsize}{!}{$
\begin{aligned}
A4(a)&=\eta \sum\limits_{h=0}^{H-1}E\Big[\Big<\nabla F(\boldsymbol{w}(t)),-\sum\limits_{i=1}^U \frac{\left|D_i\right| \beta_i}{\left|D\right|} \nabla F_i(\boldsymbol{w}_i^h(t)) \Big>\Big]\\
&=\eta \sum\limits_{h=0}^{H-1}E\Big[\Big<\nabla F(\boldsymbol{w}(t)),\big<\nabla F(\boldsymbol{w}(t))-\sum\limits_{i=1}^U \frac{\left|D_i\right| \beta_i}{\left|D\right|} \nabla F_i(\boldsymbol{w}_i^h(t)) \big>\Big>\Big] \\
&- \eta \sum\limits_{h=0}^{H-1}E\Big[\Big<\nabla F(\boldsymbol{w}(t)),\nabla F(\boldsymbol{w}(t)) \Big>\Big].\\
\end{aligned}
$}
\end{equation}

By Young's inequality, which states that for non-negative real numbers $a$ and $b$, with $p,q>1$ and $\frac{1}{p}+\frac{1}{q}=1$, we have $ab \le \frac{a^p}{p}+\frac{b^q}{q}$, and equality holds if and only if $a^p=b^q$. We can simplify Eq.~(\ref{42}) as follows:
\begin{equation}\label{43}
\resizebox{1.0\hsize}{!}{$
\begin{aligned}
&A4(a)\le \frac{\eta}{2} \sum\limits_{h=0}^{H-1} E\Bigg[ \left\| \sum\limits_{i=1}^U \frac{\left|D_i\right|}{\left|D\right|} \nabla F_i(\boldsymbol{w}_i(t))-\sum\limits_{i=1}^U \frac{\left|D_i\right| \beta_i}{\left|D\right|} \nabla F_i(\boldsymbol{w}_i^h(t)) \right\|^2 \Bigg] \\
&- \frac{\eta}{2} \sum\limits_{h=0}^{H-1} E[\left\|\nabla F(\boldsymbol{w}(t))\right\|^2 ]\\
&= \frac{\eta}{2} \sum\limits_{h=0}^{H-1} E\Bigg[ \Big\vert\Big\vert \sum\limits_{i=1}^U \frac{\beta_i\left|D_i\right|}{\left|D\right|} (\nabla F_i(\boldsymbol{w}_i(t))-\nabla F_i(\boldsymbol{w}_i^h(t)))\\
&+\sum\limits_{i=1}^U \frac{(1-\beta_i) \left|D_i\right| }{\left| D\right|} \nabla F_i(\boldsymbol{w}_i(t)) \Big\vert\Big\vert^2 \Bigg] - \frac{\eta}{2} \sum\limits_{h=0}^{H-1} [\left\|\nabla F(\boldsymbol{w}(t))\right\|^2 ].\\
\end{aligned}
$}
\end{equation}

Split the first term in inequality~(\ref{43}) into $(a1)$, $(a2)$, $(a3)$:
\begin{equation}\label{44}
\resizebox{0.9\hsize}{!}{$
\begin{aligned}
&A4(a) \le \underbrace{\frac{\eta}{2} \sum\limits_{h=0}^{H-1} E\Bigg[ \left\| \sum\limits_{i=1}^U \frac{\beta_i\left|D_i\right|}{\left|D\right|} (\nabla F_i(\boldsymbol{w}_i(t))-\nabla F_i(\boldsymbol{w}_i^h(t)))\right\|^2\Bigg]}_{(a1)} \\
&+ \underbrace{\frac{\eta}{2} \sum\limits_{h=0}^{H-1} E\Bigg[ \left\| \sum\limits_{i=1}^U \frac{(1-\beta_i) \left|D_i\right| }{\left| D\right|} \nabla F_i(\boldsymbol{w}_i(t)) \right\|^2 \Bigg]}_{(a2)} \\
&+\underbrace{\eta \sum\limits_{h=0}^{H-1}E\Bigg[ \left\| \sum\limits_{i=1}^U \frac{\beta_i\left|D_i\right|}{\left|D\right|} (\nabla F_i(\boldsymbol{w}_i(t))-\nabla F_i(\boldsymbol{w}_i^h(t))) \right\|}_{(a3)} \\
&\underbrace{\times \left\| \sum\limits_{i=1}^U \frac{(1-\beta_i) \left|D_i\right| }{\left| D\right|} \nabla F_i(\boldsymbol{w}_i(t)) \right\| \Bigg]}_{(a3)} \\
&- \frac{\eta}{2} \sum\limits_{h=0}^{H-1} E[\left\|\nabla F(\boldsymbol{w}(t))\right\|^2 ].\\
\end{aligned}
$}
\end{equation}

From the Assumption 1, the first term $(a1)$ in inequality~(\ref{44}) can be reduced to
\begin{equation}\label{45}
\resizebox{1.0\hsize}{!}{$
\begin{aligned}
&(a1)
 = \frac{\eta}{2} \sum\limits_{h=0}^{H-1} E\Bigg[ \left\| \sum\limits_{i=1}^U \frac{\beta_i\left|D_i\right|}{\left|D\right|} (\nabla F_i(\boldsymbol{w}_i(t))-\nabla F_i(\boldsymbol{w}_i^h(t)))\right\|^2\Bigg]\\
& \le \frac{\eta U L^2}{2} \sum\limits_{i=1}^U \frac{\beta_i^2\left|D_i\right|^2}{\left|D\right|^2} \sum\limits_{h=0}^{H-1} E\Big[ \left\| \nabla F_i(\boldsymbol{w}_i(t))-\nabla F_i(\boldsymbol{w}_i^h(t))\right\|^2\Big]\\
& \le \frac{\eta U L^2}{2} \sum\limits_{i=1}^U \frac{\beta_i^2\left|D_i\right|^2}{\left|D\right|^2} \sum\limits_{h=0}^{H-1} E\Big[ \left\| \boldsymbol{w}_i(t)-\boldsymbol{w}_i^h(t)\right\|^2\Big].\\
\end{aligned}
$}
\end{equation}

From the local iterative update equation for clients, we can derive the gradient update formula. The derivation process is as follows:
\begin{equation}\label{46}
\begin{aligned}
   \boldsymbol{w}_i(t)=\boldsymbol{w}_i^h(t)-\eta \sum\limits_{j=0}^{h-1}\nabla 
 F_i(\boldsymbol{w}_i^j(t)) .
\end{aligned}
\end{equation}

Substitute the gradient update Eq.~(\ref{46}) into inequality~(\ref{45}):
\begin{equation}\label{47}
\resizebox{0.9\hsize}{!}{$
\begin{aligned}
(a1)
& \le \frac{\eta U L^2}{2} \sum\limits_{i=1}^U \frac{\beta_i^2\left|D_i\right|^2}{\left|D\right|^2} \sum\limits_{h=0}^{H-1} E\Bigg[ \left\|-\eta \sum\limits_{j=0}^{h-1}\nabla F_i(\boldsymbol{w}_i^j(t))\right\|^2\Bigg]\\
& \le \frac{\eta^3 U L^2H}{2} \sum\limits_{i=1}^U \frac{\beta_i^2\left|D_i\right|^2}{\left|D\right|^2} \sum\limits_{h=0}^{H-1}\sum\limits_{j=0}^{h-1} E\Big[ \left\| \nabla F_i(\boldsymbol{w}_i^j(t))\right\|^2\Big].\\
\end{aligned}
$}
\end{equation}

Using Jensen's inequality $\left\| \sum\limits_{n=1}^N x_n\right\|^2 \le N \sum\limits_{n=1}^N \left\|x_n\right\|^2$ and property of expectations, the second term $(a2)$ in inequality~(\ref{44}) can be simplified to 
\begin{equation}\label{48}
\resizebox{0.8\hsize}{!}{$
\begin{aligned}
(a2)
&=\frac{\eta}{2} \sum\limits_{h=0}^{H-1} E\Bigg[ \left\| \sum\limits_{i=1}^U \frac{(1-\beta_i) \left|D_i\right| }{\left| D\right|} \nabla F_i(\boldsymbol{w}_i(t)) \right\|^2 \Bigg]\\
& \le \frac{\eta UH}{2} \sum\limits_{i=1}^U \frac{(1-\beta_i)^2 \left|D_i\right|^2}{\left| D\right|^2} E[ \left\| \nabla F_i(\boldsymbol{w}_i(t)) \right\|^2 ].\\
\end{aligned}
$}
\end{equation}

Substitute the Lipschitz smoothness property in Assumption 1 and Jensen's inequality into the third term $(a3)$ in inequality~(\ref{44}), we obtain:
\begin{equation}\label{49}
\resizebox{1.0\hsize}{!}{$
\begin{aligned}
(a3)
&\le \eta \sum\limits_{i=1}^U \frac{\beta_i\left|D_i\right|}{\left|D\right|} \sum\limits_{i=1}^U \frac{(1-\beta_i) \left|D_i\right| }{\left| D\right|} \\
&\quad\times \sum\limits_{h=0}^{H-1}E\Big[ \left\| (\nabla F(\boldsymbol{w}_i(t))-\nabla F(\boldsymbol{w}_i^h(t))) \right\| \cdot \left\| \nabla F_i(\boldsymbol{w}_i(t)) \right\| \Big]\\
&\le \eta L \sum\limits_{i=1}^U \frac{\beta_i\left|D_i\right|}{\left|D\right|} \sum\limits_{i=1}^U \frac{(1-\beta_i) \left|D_i\right| }{\left| D\right|} \\
&\quad\times \sum\limits_{h=0}^{H-1}E\Big[ \left\| (\boldsymbol{w}_i(t)-\boldsymbol{w}_i^h(t)) \right\| \cdot \left\| \nabla F_i(\boldsymbol{w}_i(t)) \right\| \Big] .\\
\end{aligned}
$}
\end{equation}

By substituting the gradient update Eq.~(\ref{46}) into inequality~(\ref{49}), we can further simplify $(a3)$ as follows:
\begin{equation}\label{50}
\resizebox{0.9\hsize}{!}{$
\begin{aligned}
(a3)
&\le \eta L \sum\limits_{i=1}^U \frac{\beta_i\left|D_i\right|}{\left|D\right|} \sum\limits_{i=1}^U \frac{(1-\beta_i) \left|D_i\right| }{\left| D\right|} \\
&\quad\times \sum\limits_{h=0}^{H-1}E\Bigg[ \left\| -\eta \sum\limits_{j=0}^{h-1}\nabla F_i(\boldsymbol{w}_i^j(t)) \right\| \cdot \left\| \nabla F_i(\boldsymbol{w}_i(t)) \right\| \Bigg] \\
&\le \eta^2 L \sum\limits_{i=1}^U \frac{\beta_i\left|D_i\right|}{\left|D\right|} \sum\limits_{i=1}^U \frac{(1-\beta_i) \left|D_i\right| }{\left| D\right|} \\
&\quad\times \sum\limits_{h=0}^{H-1}\sum\limits_{j=0}^{h-1}E\Big[ \left\| \nabla F_i(\boldsymbol{w}_i^j(t)) \right\| \cdot \left\| \nabla F_i(\boldsymbol{w}_i(t)) \right\| \Big] .\\
\end{aligned}
$}
\end{equation}

Substituting $(a1)$, $(a2)$, and $(a3)$ into inequality~(\ref{43}) gives
\begin{equation}\label{51}
\resizebox{0.9\hsize}{!}{$
\begin{aligned}
A4(a)
&\le \eta^3 L \sum\limits_{i=1}^U \frac{\beta_i\left|D_i\right|}{\left|D\right|} \sum\limits_{i=1}^U \frac{(1-\beta_i) \left|D_i\right| }{\left| D\right|} \\
&\quad\times\sum\limits_{h=0}^{H-1}\sum\limits_{j=0}^{h-1}E\Big[\left\| \nabla F_i(\boldsymbol{w}_i^j(t)) \right\| \cdot \left\| \nabla F_i(\boldsymbol{w}_i(t)) \right\| \Big]\\
&+ \frac{\eta^3 U L^2H}{2} \sum\limits_{i=1}^U \frac{\beta_i^2\left|D_i\right|^2}{\left|D\right|^2} \sum\limits_{h=0}^{H-1}\sum\limits_{j=0}^{h-1} E\Big[ \left\| \nabla F_i(\boldsymbol{w}_i^j(t))\right\|^2\Big]\\
&+ \frac{\eta UH}{2} \sum\limits_{i=1}^U \frac{(1-\beta_i)^2 \left|D_i\right|^2}{\left| D\right|^2} E[ \left\| \nabla F_i(\boldsymbol{w}_i(t)) \right\|^2 ]\\
&- \frac{\eta}{2} \sum\limits_{h=0}^{H-1} E[\left\|\nabla F(\boldsymbol{w}(t))\right\|^2 ].
\end{aligned}
$}
\end{equation}

Inequality~(\ref{41}) can be converted to the following expression:
\begin{equation}\label{52}
\resizebox{0.9\hsize}{!}{$
\begin{aligned}
&E[F(\boldsymbol{w}(t+1))]\le E[F(\boldsymbol{w}(t))]+\eta^2 L \sum\limits_{i=1}^U \frac{\beta_i\left|D_i\right|}{\left|D\right|} \sum\limits_{i=1}^U \frac{(1-\beta_i) \left|D_i\right| }{\left| D\right|} \\
&\times\sum\limits_{h=0}^{H-1}\sum\limits_{j=0}^{h-1} E\Big[\left\|\nabla F_i(\boldsymbol{w}_i^j(t)) \right\| \cdot \left\| \nabla F_i(\boldsymbol{w}_i(t)) \right\| \Big]\\
&+\frac{\eta^3 U L^2H}{2} \sum\limits_{i=1}^U \frac{\beta_i^2\left|D_i\right|^2}{\left|D\right|^2} \sum\limits_{h=0}^{H-1}\sum\limits_{j=0}^{h-1} E\Big[ \left\| \nabla F_i(\boldsymbol{w}_i^j(t))\right\|^2\Big]\\ 
&+\frac{\eta^2 LUH}{2} \sum\limits_{i=1}^U \frac{\beta_i^2 \left|D_i\right|^2}{\left|D\right|^2} \sum\limits_{h=0}^{H-1} E\Big[ \left\| \nabla F_i(\boldsymbol{w}_i^h(t))  \right\|^2 \Big]\\
&+ \frac{\eta UH}{2} \sum\limits_{i=1}^U \frac{(1-\beta_i)^2 \left|D_i\right|^2}{\left| D\right|^2} E[ \left\| \nabla F_i(\boldsymbol{w}_i(t)) \right\|^2 ]\\
&- \frac{\eta}{2} \sum\limits_{h=0}^{H-1} E[\left\|\nabla F(w(t))\right\|^2 ]\\
\end{aligned}
$}
\end{equation}

By summing over all $T$ communication rounds and rearranging the terms in inequality~(\ref{52}), we can obtain:
\begin{equation}\label{53}
\resizebox{0.9\hsize}{!}{$
\begin{aligned}
&\frac{1}{T}\sum\limits_{t=1}^T E[\left\|F(\boldsymbol{w}(t+1))\right\|^2] \le \frac{2 \eta L}{HT} \sum\limits_{i=1}^U \frac{\beta_i\left|D_i\right|}{\left|D\right|} \sum\limits_{i=1}^U \frac{(1-\beta_i) \left|D_i\right| }{\left| D\right|} \\
&\times\sum\limits_{h=0}^{H-1}\sum\limits_{j=0}^{h-1} E\Big[\left\|\nabla F_i(\boldsymbol{w}_i^j(t)) \right\| \cdot \left\| \nabla F_i(\boldsymbol{w}_i(t)) \right\| \Big]\\
& +\frac{\eta^2 U L}{T} \sum\limits_{i=1}^U \frac{\beta_i^2\left|D_i\right|^2}{\left|D\right|^2} \sum\limits_{t=1}^{T} \sum\limits_{h=0}^{H-1}\sum\limits_{j=0}^{h-1} E\Big[ \left\| \nabla F_i(\boldsymbol{w}_i^j(t))\right\|^2\Big]\\
&+\frac{\eta LU}{T} \sum\limits_{i=1}^U \frac{\beta_i^2\left|D_i\right|^2}{\left|D\right|^2} \sum\limits_{t=1}^{T}\sum\limits_{h=0}^{H-1} E\Big[ \left\| \nabla F_i(\boldsymbol{w}_i^h(t))  \right\|^2 \Big]\\
&+\frac{U}{T}\sum\limits_{i=1}^U \frac{(1-\beta_i)^2 \left|D_i\right|^2}{\left| D\right|^2} \sum\limits_{t=1}^{T}E[ \left\| \nabla F_i(\boldsymbol{w}_i(t)) \right\|^2 ] \\
& + \frac{2 E[F(\boldsymbol{w}(1))-F(\boldsymbol{w}^*)]}{\eta HT}\\
\end{aligned}
$}
\end{equation}


If we substitute Assumption 2 and further simplify, then inequality~(\ref{53}) can be converted to
\begin{equation}\label{55}
\begin{aligned}
&\frac{1}{T}\sum\limits_{t=1}^T E[\left\|F(\boldsymbol{w}(t+1))\right\|^2] \\
&\le \frac{2 E[F(\boldsymbol{w}(1))-F(\boldsymbol{w}^*)]}{\eta HT}+UG^2\sum\limits_{i=1}^U \frac{(1-\beta_i)^2 \left|D_i\right|^2}{\left| D\right|^2}\\
&+  2\eta L HG^2 \sum\limits_{i=1}^U \frac{\beta_i\left|D_i\right|}{\left|D\right|} \sum\limits_{i=1}^U \frac{(1-\beta_i) \left|D_i\right| }{\left| D\right|}\\
&+\eta ULH G^2*(\eta H + 1) \sum\limits_{i=1}^U \frac{\beta_i^2\left|D_i\right|^2}{\left|D\right|^2}.\\
\end{aligned}
\end{equation}

This completes the proof.

}

\bibliographystyle{IEEEtran}
\bibliography{bib2}



 





\end{document}